\documentstyle[twoside,fleqn,espcrc2]{article}

\newcommand{\AmS}{{\protect\the\textfont2
  A\kern-.1667em\lower.5ex\hbox{M}\kern-.125emS}}
\newcommand{\Tr}[1]{\,{\rm Tr}\,#1\,}
\title{Chiral symmetry in lattice QCD}

\author{A.A.Slavnov \address{Steklov Mathematical Institute, Russian
        Academy of Sciences, Gubkina st. 8, GSP-1, 117966,Moscow, Russia}%
       \thanks{supported in part by RBRF grant 99-01-00190 and INTAS-9670}}
 \begin{document}
\def\fmn{{\cal F}_{\mu\nu}}
\def\am{{\cal A}_{\mu}}
\def\ama{\am^{a}}
\def\qm{q_{\mu}}
\def\qma{\qm^{a}}
\def\sn{S_{n}^{\Lambda}}
\def\kongo{\frac{1}{2}\int\!\int\frac{\delta^{2}\sn}{\delta\ama(x)\delta{\cal
A}_{\nu}^{b}(y)}\qma(x) q_{\nu}^{b}(y)dxdy}
\def\kon{\Re_{n} (\am,\qm)}
\def\gft{\frac{1}{4\alpha}{\bf
tr}\{f_{n}(\frac{\Box}{\Lambda^2})\,\partial_{\mu}\am \}^{2}}
\def\gftq{\frac{1}{4\alpha}{\bf tr}\{f_n
(\frac{\Box}{\Lambda^2})\,\partial_{\mu}\qm \}^2} \def\za{{\cal
Z}_{\alpha}} \def\zo{{\cal Z}_{0}} \def\zinv{{\cal Z}_{\mbox{inv}}}
\def\gda{\det{\cal M}}
\def\dd{\prod_{x}} \def\pdm{\partial_{\mu}} \def\nm{\nabla_{\mu}}
\def\Fn{F_{n}\Bigl(\frac{\nabla^2}{\Lambda^2}\Bigr)}
\def\bFn{\frac{1}{2\beta} \int [\Fn \nm \qm]^{2}dx}
\def\cj{c_{\jmath}}
\def\dnm{det (\nabla^2-M^2)}
\def\dnmj{det^{\cj} (\nabla^2 - M_{\jmath}^2)}
\def\dqm{det_{n}^{-1/2}Q_{\beta}(\am;M^2;F_{n})}
\def\dqmj{det_{n}^{-\cj /2}Q_{\beta}(\am;M_{\jmath}^{2};F_{n})}
\def\zlm{{\cal Z}^{n}_{\Lambda,M^2}[J]}
\def\hact{\Bigl[\frac{\delta S_0}{\delta {\cal A}_{\rho}(z)}+\nabla_{\rho}
\Box^2\partial {\cal A}(z)\Bigr]}
\def\soa{\frac{\delta^2S_0}{\delta \am (x)\delta {\cal A}_{\rho}(z)}}
\def\opdet{\Bigl[\soa+\nabla_{\rho}\Box^2\pdm \delta(x-z)\Bigr]}
\def\detdet{det^{1/2}\Bigl[\Box^2\nabla^2\Box^2+\alpha^{-1}\Lambda^{10}\Bigr]}
\def\finv{ F\Bigl(\frac{\nabla^2}{\Lambda^2}\Bigr)}
\def\klio{det^{1/2}\Bigl[\nabla^2+\alpha^{-1}\Box^{-4}\Lambda^{10}\Bigr]}
\def\cov{\frac{1}{\Lambda}(\nm\nabla_\rho-\nabla^2g_{\mu\rho})\delta(x-z)}

\begin{abstract}
A chiral invariant effective action for lattice QCD is proposed. Its
connection to the multifermion model is established. A possibility
of using this action for computer simulations is discussed.  \end{abstract}

\maketitle

\section{Introduction.}

Approximate chiral symmetry is known to play an essential role in QCD.
Hence a faithful transcription of chiral symmetry in lattice
regularization of QCD is of prime importance. Recently an important
progress was achieved in this direction (see
e.g.  \cite{K}, \cite{Sl.Fr}, \cite{N.N}, \cite{Sh}, \cite{Has}, \cite{N},
\cite{Lu}, \cite{Sl.1}). In particular it was understood that
Nielsen-Ninomiya "no-go" theorem \cite{Ni.Ni} may be bypassed by avoiding
some of its assumptions. Nevertheless all models used so far for numerical
simulations do not posess exact chiral invariance.

The most important problem at present is to formulate a lattice version of
QCD which allows efficient numerical simulations producing  a "minimal"
breaking of chiral symmetry. By a "minimal" breaking I mean that the model
does not require chiral noninvariant counterterms in the continuum limit
and for a finite lattice spacing symmetry breaking effects can be done
negligible in a reasonable computer time.

In my paper \cite{Sl.1} an effective action for lattice
QCD, based on the idea of multifield formalism \cite{Sl.Fr.1},
was proposed, which provides exponential suppression of chirality
breaking effects.  In the present talk
I show that this model may be described by a manifestly chiral invariant
single field action. This action is formally nonlocal but possible
nonlocal effects are suppressed.

\section{The model.}

The effective action for lattice QCD may be taken in the form
\begin{eqnarray}
I= I_W+ \nonumber\\ \sum_{x, \mu}
\bar{\psi}(x) \hat{D} \coth(\frac{ \pi}{2}|(i \hat{D})^{1/2}D^{-2}
(i \hat{D})^{1/2})| \psi(x), \label{1} \end{eqnarray}
where $I_W$ is the usual Wilson action for gluons,
\begin{eqnarray}
\hat{D}= \gamma_{\mu}(D_{\mu}+ D^*_{\mu}) \nonumber\\
D^2= \kappa aD^*_{\mu}D_{\mu}
\label{1a}
\end{eqnarray}
 and $D_{\mu}$ is the lattice covariant derivative
\begin{equation} D_{\mu} \psi= \frac{1}{a}[U_{\mu}(x) \psi(x+a_{\mu})-
\psi(x)] \label{2}\end{equation} In the formal continuum limit $a
\rightarrow 0$,  the path integral of the exponent of this
action obviously reproduces the usual QCD partition function.

The action (\ref{1}) is manifestly chiral invariant, but formally
nonlocal.  We shall show however that this nonlocality is harmless as
nonlocal effects disappear in the continuum limit and are suppressed for a
finite lattice spacing. No spectrum doubling is generated by the
action (\ref{1}).

However this action is too complicated to be used for computer
simulations. Luckily, the corresponding path integral may be also written
in terms of a local multifield action which is suitable for numerical
simulations. This local action was introduced before in the paper
\cite{Sl.1}
  \begin{equation} \tilde{I}=I_W+\sum_{n=-
\infty, n \neq 0}^{+ \infty} \sum_{x} \bar{\psi}^n(x)[\hat{D}-nD^2]
\psi^n(x) \label{3} \end{equation} Here the spinor fields $ \psi^n$ have
the Grassmaninan parity $(-1)^{n+1}$.  Integrating the exponent of the
action (\ref{3}) over $\bar{\psi}_n, \psi_n$, we get \begin{eqnarray} Z=
 \prod_{n=- \infty, n \neq 0}^{+ \infty}
 \det(\hat{D}-nD^2)^{(-1)^n} \nonumber\\ = \exp
 \{\sum_n (-1)^n \Tr \ln[\hat{D}-nD^2] \} \label{4}
 \end{eqnarray}
The exponent in the eq.(\ref{4}) may be presented in the form
\begin{eqnarray}
\sum_n (-1)^n \Tr \ln[ \hat{D}-nD^2]= \nonumber\\
\lim_{ \Lambda \rightarrow \infty} \{ \Tr[- \hat{D} \int_1^ \Lambda
d \alpha \sum_n(-1)^n(\alpha \hat{D}- \nonumber\\ nD^2)^{-1}]+
 \sum_n(-1)^n \Tr \ln( \Lambda \hat{D}-nD^2) \}
 \label{5}
 \end{eqnarray}
 The limit of the second term at the r.h.s. is equal to
 $$
 \Tr \ln( \hat{D})+ const
 $$
and in the first term we use the following relation
 \begin{eqnarray} \Tr[ \hat{D}( \alpha
\hat{D}-nD^2)^{-1}]= \nonumber\\ \Tr[ \alpha \hat{I}-n \hat{D}^{-1/2}D^2
\hat{D}^{-1/2}]^{-1}= \sum_k( \alpha+inB_k]^{-1} \label{6} \end{eqnarray}
where $B_k$ are eigenvalues of the operator \begin{equation} B=(i
\hat{D})^{-1/2}D^2 (i \hat{D})^{-1/2} \label{6a} \end{equation} Using this
representation one can perform the summation over $n$ in the eq.(\ref{4})
  explicitely with the result:  \begin{eqnarray} \lim_{\Lambda \rightarrow
  \infty} \int_1^{\Lambda}d \alpha \sum_k[ \frac{ \pi |B_k^{-1}|}{\sinh(
 \pi \alpha |B_k^{-1}|)}+ \frac{1}{\alpha}]= \nonumber\\ \lim_{\Lambda
  \rightarrow \infty} \sum_k[ \ln \coth(\frac{\pi \Lambda |B_k^{-1}|}{2})-
  \nonumber\\ \ln \coth(\frac{\pi |B_k^{-1}|}{2})] \label{7}
 \end{eqnarray} where a nonessential constant was omitted.

So we proved that
 \begin{eqnarray}
 \prod_{n=- \infty, n \neq 0}^{\infty} \det( \sum_{\mu}[
 \gamma_{\mu}(D_{\mu}+D^*_{\mu})- \nonumber\\ na \kappa
D^*_{\mu}D_{\mu}])^{{-1}^n}= \nonumber\\ \det[ \hat{D}
\coth(\frac{\pi}{2}[|(i \hat{D})^{1/2}D^{-2}(i \hat{D})^{1/2}|)]
\label{8} \end{eqnarray} which coincides with the expression obtained by
integration the exponent of the action (\ref{1}) over $\bar{\psi}, \psi$.

 We have two alternative representations for the determinant in the r.h.s.
 of equation (\ref{8}). The first one is given by the integral of the
 exponent of the manifestly chiral invariant but formally nonlocal action
 (\ref{1}), and the second one is the integral of the multifield action
 (\ref{3}). The last action is formally local, but includes infinite
 series of auxilliary fields. One may suspect that this infinite summation
 may generate some nonlocal effects. Such a possibility indeed exists,
 however due to exponential convergence of the series in the eq.(\ref{5})
 one may cut the series by some finite number $N$ and for $N$ big enough
 the correction term is small. For a finte $N$ the action (\ref{3}) is
 local. It proves that both representations define an "almost" local
 theory, which becomes exactly local in the continuum limit. The
 representation in terms of the effective action (\ref{1}) makes chiral
 symmetry manifest. Possible counterterms must respect chiral
 symmetry. In principle these counterterms could be nonlocal , as it
 happens in SLAC model \cite{K.S}, however the alternative
 representation of quark determinant as the path integral of the local
 multifield action (\ref{3}) shows that it does not happen. The explicit
 proof of the exponential suppression of chirality breaking effects in
 perturbation theory for the model described by the multifield
 action(\ref{3}) was given in the paper \cite{Sl.1}.

The nonlocal action (\ref{1}) is not suitable for numerical simulations.
In the alternative multifield formulation one has to calculate the product
of the usual Wilson fermion determinants. However the effective action
(\ref{3}) includes an infinite number of fields and to make simulations
one has to cut the series.  The crucial question is how sensitive is the
result to the cutting the series by some finite $N$. The estimate given in
the paper \cite{Sl.1} in the framework of perturbation theory shows that
for external momenta $q$ satisfying the relation $|qa| \ll 1$ a small
number of auxilliary fields is sufficient.  This conclusion was also
supported by numerical simulations in two-dimensional models.  However
only real four-dimensional simulations may check the efficiency of the
method.  The estimates show that the convergence may slow down for small
eigenvalues of the operator $B$ (eq.(\ref{6a})). A nonperturbative study
of the spectrum of this operator would be important for the estimate of
convergence rate.

 \section{Discussion}

 It was shown in the previous section that the lattice QCD quark
determinant may be presented in two alternative forms. In one form it
is given by the path integral of the exponent of the chiral
invariant but formally nonlocal effective action (\ref{1}). Another form
is given by the path integral of the local multifield action (\ref{3}),
which is not manifestly chiral invariant. We proved that both these forms
represent the same quark determinant which enjoys therefore both locality
and chiral invariance. In the framework of perturbation theory it was
checked before that our model in the continuum limit reproduces exactly
the quark determinant of massless QCD without any chiral noninvariant
counterterms.  For a finite lattice spacing chirality breaking corrections
are exponentially small $$ \sim O( \exp \{-( \kappa \epsilon)^{-1} \}),
\quad \epsilon \sim |q|a $$, where $q$ is a maximal external momentum.

Finite quark masses may be easily incorporated into our scheme. One should
modify the effective action (\ref{1}) as follows
\begin{eqnarray}
I= I_W+ \nonumber\\ \sum_{x, \mu}
\bar{\psi}(x) \hat{D} \coth(\frac{ \pi}{2}|(i \hat{D}+m)^{1/2}D^{-2}
\times \nonumber\\ (i \hat{D}+m)^{1/2}|) \psi(x) \label{9}
 \end{eqnarray} Of course in this case the chiral invariance is
 explicitely broken by the bare quark mass.

{\bf Acknowledgements.}\\ I am grateful to the organizers of the
Conference "Lattice-00" in particular Apoorva Patel for
hospitality and financial support.

\end{document}